# A new model for the implementation of positive and negative emotion recognition

Jennifer Sorinas[a, b*], Juan C. Fernandez-Troyano, Mikel Val-Calvo[b], Jose Manuel Ferrández[b] and Eduardo Fernandez[a]

[a] Institute of Bioengineering, University Miguel Hernandez and CIBER BBN, Elche, 03202, Spain
[b] Department of Electronics and Computer technology, University of Cartagena, Cartagena, 30202, Spain

**ABSTRACT**

The large range of potential applications, not only for patients but also for healthy people, that could be achieved by affective BCI (aBCI) makes more latent the necessity of finding a commonly accepted protocol for real-time EEG-based emotion recognition. Based on wavelet package for spectral feature extraction, attending to the nature of the EEG signal, we have specified some of the main parameters needed for the implementation of robust positive and negative emotion classification. 12 seconds has resulted as the most appropriate sliding window size; from that, a set of 20 target frequency-location variables have been proposed as the most relevant features that carry the emotional information. Lastly, QDA and KNN classifiers and population rating criterion for stimuli labeling have been suggested as the most suitable approaches for EEG-base emotion recognition. The proposed model reached a mean accuracy of 98% (s.d. 1.4) and 98.96% (s.d. 1.28) in a subject-dependent approach for QDA and KNN classifier, respectively. This new model represents a step forward towards real-time classification. Moreover, although results were not conclusive, new insights regarding subject-independent approximation have been discussed.

**Keywords**: aBCI, EEG, emotion recognition, wavelets

1. **Introduction**

Brain Computer Interface (BCI) implies the control of a device without the intervention of the peripheral neural system by using signals coming from the brain, monitored through neuroimaging tools; with the aim of helping patients to compensate loss of motor function [1]. In recent years, BCI has led research towards the necessity of improving human-computer communication in order to achieve more precise user experience, more efficient computer work and to enlarge applications towards healthy people. The key to achieving this goal is the detection and recognition of user' emotions; offer a response to instant emotions without the need to out express them. This novel emerging field, known as aBCI, is based on systems that measure signals related to affective states from the peripheral and central nervous systems, with the intention to adapt the human-computer interaction. aBCI would provide more life quality to BCI user' patients, and would let to a new field of therapies, modality called neurofeedback [2], through human-computer interactions that could aid on the treatment of some mental disorders as for example, autism [3], Attention Deficit Hyperactivity Disorder (ADHD) [4], depression [5], schizophrenia [6], and epilepsy [7]. Neurofeedback is not only focus on the treatment or evaluation of the status of a disease [8], it may be also suitable for improving cognitive performance [9]; and be used on entertainment and gaming applications [10]. Therefore, aBCI is more than a potential advantage for the treatment of certain disorders; its practice may also be open to healthy users. To that end, real-time BCI systems are needed; and consequently, algorithms for every stage of the BCI process (emotion elicitation, feature extraction, feature selection and classification) should be correctly specified and implemented. There is a high diversity of methodologies to carry out each level of the process, what increments variability between studies and makes difficult the existence of a commonly accepted model that could ensure accurate and reproducible results. Although progress have been made, the difficulty increases facing real-time applications, where the speed of the process is a crucial factor in algorithm selection. Moreover, currently BCI systems, yield on subject-dependent (SD) classification models where the classifier should be trained for each user prior to any sorting and personalized feature selection is needed. Due to the time consuming cost that generates the training of every single subject, the formula for the ideal subject-independent (SI) approach have been searched. However, the design of accurate models is still a great challenge.

When studying real-time emotion detection, properly emotional stimulation is crucial. The stimuli should reflect accurately the desired process regarding the model of emotion [11] [12]; and being ecological in terms of real world approximation, where information comes from all senses; therefore, dynamic stimuli are preferred [13] [14]. In aBCI real-time applications, the EEG technique has the prominence among the neuroimaging methods, due to its high temporal resolution, safety non-invasive use,

∗ Corresponding author.
E-mail addresses: jennifersorinas@gmail.com (*Jennifer Sorinas*), jcftroyano@gmail.com (*J.C. Fernandez-Troyano*), mval33@alumno.uned.es (*Mikel Val-Calvo*), jm.ferrandez@upct.es (*J.M. Ferrández*), e.fernandez@umh.es (*Eduardo Fernandez*)

relatively low-cost, portability and therefore, accessibility to all levels from research to clinics passing through private companies [15]. For all these advantageous characteristics, research in the recognition of human emotions has focused on the implementation of this technique. The different bandwidths of the EEG spectrum have been associated with multitude of cognitive functions as well as affective states [16]. Traditionally, linear methods, which are effective in terms of computational time cost, were used for power spectral feature extraction. However, the EEG signal is nonstationary, i.e. it doesn't obey constant mean and standard deviation over time; and nonlinear, it doesn't follow homogeneity and superposition principles [17]; hence, nonlinear extraction methods, that respect the nature of the signal, are more eligible to achieve robust classification results, being the wavelet transform the most recommended [18]. The most significant specification of feature extraction is the size of the sliding window to ensure suitable time frequency resolution in all frequency ranges [19]. On the other hand, one of the major challenges in BCI applications is the selection of the relevant features that contain the neuronal information relative to the process of study. The specific components of the EEG signal that encode the emotional processes itself, are going to be localized in space, particular cortical locations, and constrained to a scale of frequency, explicit bandwidths. The identification and selection of the informative features could generate four benefits. First, the existence of less irrelevant and redundant data reduces the probability of modelling noise instead of the studied process, hence, reducing the overfitting tendency. Second, in some cases, classification accuracy could improve. Third, classifier training time is reduced and so, time consuming cost. Finally, a high number of sensors means less comfort for the user, higher system complexity, increased cost and more classification challenge due to the high-dimensional feature space [20]. Eventually, the selection of the classifier that best adjust to the bias-valence ratio is desirable. High-variance learning methods, have enough flexibility to represent complex data distributions even with the presence of noisy variations, so they could potentially model unrepresentative properties. In contrast, high-bias algorithms, are not suitable for representing complex data distributions, leading to poor models, when failing to grasp the complex nature of the training samples. Although, during recent years more classifiers have become to be applied in BCI systems [21], linear classifiers continue to be the most popular for real-time applications, particularly LDA [22] and support vector machines (SVM) [23], without achieving high classification performances but being efficient in terms of time cost.

The large number of existing methodologies when implementing BCI stages, makes decidedly difficult the comparison between studies; ergo, the existence of a commonly accepted protocol for EEG-based emotion recognition. The present work, performed offline and in continuity with our previous work [24], corresponds to a feasibility study for the implementation of real-time positive and negative emotion classification based on EEG. We have used the same self-done database used in previous works [24] [25], divided into positive and negative emotional content video clips, based on the biphasic emotional model [26]. The audiovisual excerpts are not popular, in order to avoid familiarity [27], and produce a sustained emotion during their whole duration. We have addressed the problem of the sliding window size and the specification of the informative frequency-location features by using nonlinear feature extraction methodology, wavelet transform, expecting better results in comparison with linear methods. We have assessed the performance of a battery of sorters, linear and quadratic discriminant analysis (LDA and QDA, respectively), support vector machines (SVM), nearest neighbors (NN), naive Bayes (NB) and ensemble methods, in order to select the most suitable one for real-time EEG-based emotion recognition. We have focus on a SD approach, to specify the mentioned parameters for a common automatized SD real-time classification, in order to facilitate aBCI processes and reduce system complexity and training time. In addition, a first approximation for the search for an SI model is briefly discussed. Different stimuli label criteria were also evaluated. In the end, we have proposed a new model for achieving SD positive and negative emotion recognition and new insights for the searching of the desired SI model.

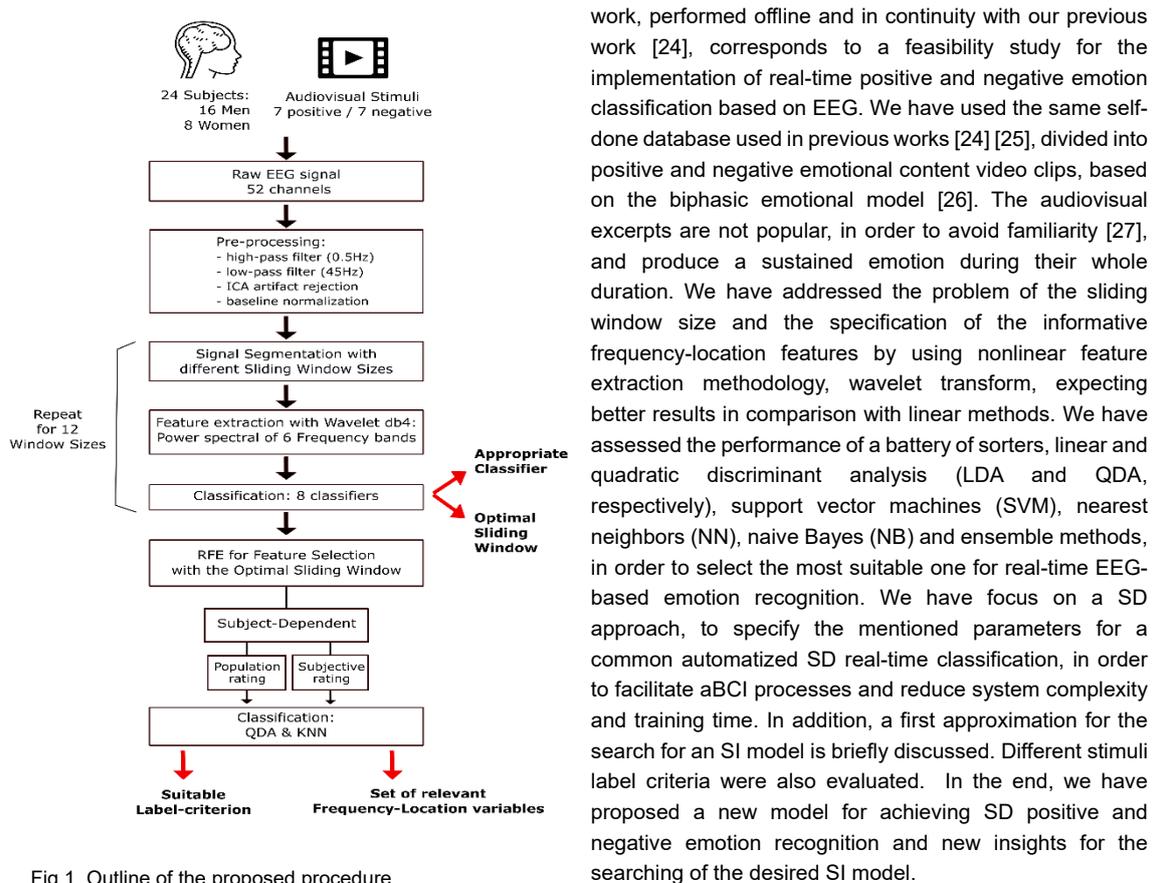

Fig.1. Outline of the proposed procedure

## 2. Material and methods

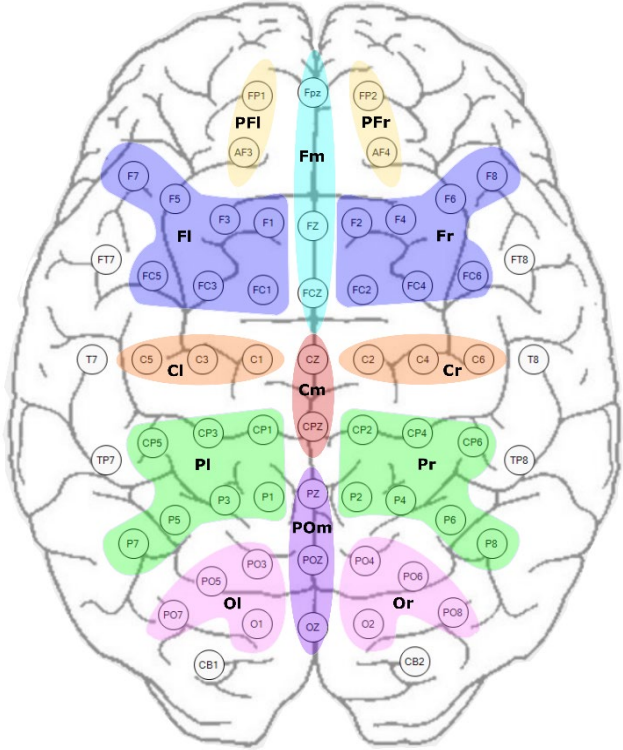

Fig 2. Distribution of the sets of electrodes in 13 functional regions. PFr: Pre-Frontal right, dark yellow. PFl: Pre-Frontal left, light yellow. Fl = Frontal left, Fr = Frontal right, Fm = Frontal midline, Cm = Central midline, Cl = Central left, Cr = Central right, Pl = Parietal left, Pr = Parietal right, Ol = Occipital left, Or = Occipital right, POm = Parietal-occipital midline.

### 2.1. Experimental Procedure

We have a sample size of 24 subjects (mean age: 23.12; range: 19–37; sixteen men and eight women). All of them were right-handed, without personal history of psychiatric or neurological disorders, without currently medication and normal or corrected vision and audition. The participants provided their written consent, supervised by the Ethics Committee of the University Miguel Hernandez.

Emotion elicitation was performed through audiovisual stimulation by 14 video clips selected from internet with durations between 43 and 78 seconds. 7 videos were labelled as positive emotional content, for example they represent natural landscapes or cartoons; on the other hand, the other 7 remaining videos that contained scenes of violence or horror stories, were classified into the negative emotional category. The audiovisual excerpts were presented in a random mode and counterbalanced for each subject, in alternation with a 30 seconds black screen. During the black screen periods after each video, the participants were instructed to rate the emotional content of the former clip into valence, ranging from 9 (very pleasant) to 1 (very unpleasant), and arousal, ranging from 9 (very nervous or goad) to 1 (very relaxed or calm), dimensions. The whole methodology is summarized in Fig. 1.

### 2.2. Signal Acquisition and Pre-processing

EEG data was recorder through a cap-mounted 64 Ag-AgCl electrodes according to the International 10/10 System [28], and amplified by the NeuroScan SynAmps EEG amplifier (Compumedics, Charlotte, NC, USA). The recordings, acquired with 1000Hz of sample rate, were re-referenced to a Common Average Reference (CAR) and filtered using a high-pass and low-pass filters, 0.5Hz and 45Hz respectively, by Curry 7 software (Compumedics, Charlotte, NC, USA). 52 electrodes from the total of 62 were finally used for the study. The Matlab toolbox EEGLAB [29] was used to reject the artifacts corresponding to muscle noise, eye-blinking and heart rate, selected by visual inspection based on the Independent Component Analysis (ICA) [30]. See previous work for more details related with the data collection and pre-processing [24].

In order to emphasize the differences between the EEG signals during rest period (initial black mute screen, also called baseline), and audiovisual stimulation; the signal (s) of each electrode was z-scored, as in (1), with respect to the baseline (b). We used excerpts of the same amount of time for both stimuli and baseline, corresponding to the first 28 seconds of recording but removing the first second to avoid premier sensory impact.

$$Zscore = \frac{s(t) - \overline{b}}{\sigma(b)} \quad (1)$$

The preprocessed and standardized EEG data of each video clip were segmented into different sliding windows ranging from 1 to 12 seconds with a step of 1 second. The different segments of window sizes (t) were based on the work carried out by Candra et al. [19], were they concluded that the suitable window size for emotion recognition based on EEG data might be between 3s and 12s. Every resulting segment was considered as a trial for further analysis.

### 2.3. Feature Extraction

In order to capture and analyze the nonlinear nature of EEG signals, a non-parametric feature extraction method based on multi-resolution analysis of wavelet transform was used. The EEG channels were arranged into 13 functional assemblies, see Fig. 2. The spectral power of a predefined set of frequency bands (delta, theta, alpha, beta1, beta2, gamma), based on accepted ranges of brain oscillations, was extracted for every set of electrodes. Level 8 wavelet packets decomposition was performed in order to have enough resolution over the frequency range, and Daubechies order 4 (db4) mother wavelet was used [31]. Db 4 has resulted as the most suitable mother wavelet for assessing human emotions from EEG signal by several authors [32] [33] [34] [35]. Wavelet power spectrum estimates were computed, obtaining the EEG signal information in the time-frequency domain for every trial. The frequency resolution was approximately 2Hz and time was equal to the duration of the sliding-window. Finally, to only consider the information in the frequency domain, each power spectrum estimate was averaged over time in the specified bandwidths.

*2.4. Optimal Sliding Window Estimation*

A set of classifiers was used to analyze the discrimination of EEG signal data distributions in different sliding windows, ranging from 1 to 12 seconds. Power spectral frequency for every bandwidth extracted at each window time and at every set of electrodes, was used as a classifier input. Data were standardized by removing the mean and scaling to unit variance. The classifiers evaluated were: LDA and QDA discriminant analysis methods; SVM with linear and nonlinear (radial basis function kernel) kernels; k-nearest neighbors (KNN) classification with 5 neighbors; the Gaussian Naive Bayes algorithm (GNB); and the Gradient Boosting (GB) with 10 estimators, and Random Forest (RF) decision tree with 100 estimators, both ensemble methods based on boosting and averaging methods, respectively. Classification stage was performed assuming the SD design. An iterative process with an 8-fold cross validation was made. The data set was split randomly into 8 folds ensuring that the balance between classes was maintained. On each iteration one of the folds was fixed as the validation set and the rest as the training set. F1 score (2) and accuracy (3) statistical metrics were computed for every iteration based on the test set classification results. Mean F1 score and accuracy were obtained for every subject and classifier. We have chosen the f1 score, because it penalizes those results that have a clear decantation towards one classification category at the expenses of the other, i.e. overfitting. Best sliding window and optimal classifier were specified based on classification performance for the sample ensemble, i.e. mean results across subjects. Scikit-learn and pandas libraries for python for machine learning and data analysis were used for the whole computing process [36], [37].

One sample Kolmogorov-Smirnov test was applied to each window size to assess the normal distribution of the data [38]. The Friedman's test [39], followed-up by multiple comparison test [40], was applied to evaluate the statistical differences between the different sliding window sizes. The same statistical procedure was this time applied for the comparison of the performance of the classifiers, on the best sliding window size. The p-value below which the differences are considered statistically significant was 0.05.

$$F1_{score} = \frac{2*\text{precision}*\text{recall}}{(\text{precision}+\text{recall})} \quad (2)$$

$$Acc = \frac{TP+TN}{TP+TN+FP+FN} \quad (3)$$

Where Acc means accuracy, TP = true positives, TN = true negatives, FP = false positives, and FN = false negatives.

*2.5. Feature Selection*

Once that the optimal sliding window was estimated, the feature selection method known as recursive feature elimination (RFE), was applied with the aim of optimizing classification performance by dimensionality reduction. The process was carried out assuming SD classification model and two stimuli label criterion, the first one based on the mean valence population ratings (PR) of every video for the 24 subjects, and a second criteria were the trial label was personalized for every subject attending to the individual subjective ratings (SR). For model fitting, SVM with linear kernel was used and a subset of the best 20 features were selected for every subject at both the PR and SR label criterion. From the total 480 selected frequency-location variables, the 20 most repeated features across subjects were selected for final classification stage.

On the other hand, simulated annealing (SA) [41] method, with 300 interactions, was implemented for the best sliding window size and all the tested classifiers, assuming the SI model for global optimization. Also, an ensemble of features was selected for every classifier and used for final classification step.

*2.6. Classification*

Final classification stage was performed at the best sliding window, with the optimal classifier and taking into account different sets of features but always based on those selected on previous dimensionality reduction step. For the SD-PR group, 5 classification approaches were evaluated: (1) using the 5 most common features selected across subjects at the RFE analysis; (2) using the 10 most common features; (3) using the 15 most common features; (4) using the 20 most common features; and (5) using 20 subject-specific features. For the SD-SR group classification was carried out based on the 20 subject-specific features. The classification was performed following the same methodology applied in the estimation of the optimal sliding window, section 2.4, with 8-fold cross validation, obtaining mean f1 scores and accuracies for every subject.

Lastly, for de SI group, using the PR criterion, either the 20 common SD-PR features selected with the RFE analysis at section 2.5, and the features selected with the SA method at the SI approach, were used for final classification testing again the collection of classifiers. In this case, classification was performed using leave-one-subject-out cross validation, where 24 iterations were performed fixing one subject as the test set and the other 23 as the training set per iteration. F1 score and total accuracy statistical metrics were obtained for the population sample.

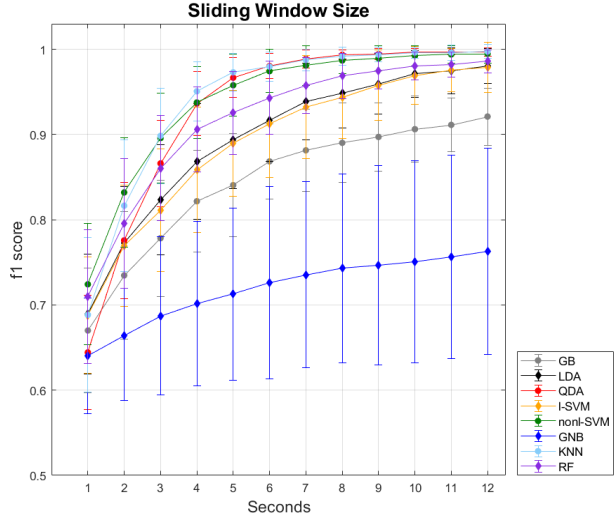

Figure 3. Mean classification performance across subjects of the eight tested classifiers at the twelve evaluated window sizes.

The statistical analysis comparing the different classification approximations was carried out in the same way as in section 2.4.

## 3. Results

### 3.1. Stimuli Subjective Ratings

By looking at the mean valence and arousal values that the 24 subjects gave to the video clips, PR criterion, the emotional content of the videos can be clearly separated into positive and negative categories in the valence scale. The mean values obtained at the valence scale were 7.51 (standard deviation or std 1.6) for the positive emotional group and 2.91 (std 0.98) for the negative group. At the arousal scale the positive emotional content videos obtained a mean of 3.76 (std 1.62); the group of negative videos obtained a mean value of 5.47 (std 1.35). Unlike on the valence scale, the differences at the arousal scale were not that clear. However, our interest resides on the valence scale that represents the approach/withdrawal motivational systems. On the other hand, if we look deep into the subjective ratings, 14 subjects rated all the videos as expected taking into account the storyline of themselves. 10 subjects, corresponding to subject 4, 5, 6, 8, 9, 17, 18, 19, 20, 21, missed their ratings in at least 2 video clips, providing positive valence values to negative content videos and/or vice versa.

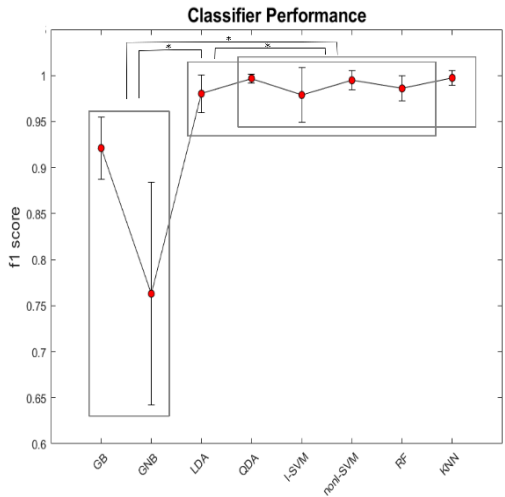

Figure 4. Statistical differences between the evaluated classifiers. The rectangles show the different groups between which there are significantly differences with p-values lower than 0.05 (*).

### 3.2. Suitable Window Size

We have focused on the SD classification model to assess the optimal window size for the extraction of the power spectral information that best reflects the valence dimension of the emotional process. The mean f1 score for the 24 subjects was computed for every window size at each of the eight tested classifiers (Fig. 3). Two preliminary conclusions could be extracted. (1) Not all the classifiers presented the same performance, being the GNB the one that obtained worse results. Moreover, non-linear classifiers performed better than their linear versions (e.g. LDA and SVM). (2) There was a generalized positive correlation between the size of the sliding window and the classification performance. With the intention of statistically specified the best sliding window size for wavelet based feature extraction regarding emotions, we have compared the mean classification performances of the classifiers at each time window. It resulted that, the 12 seconds time window showed best classification performance, and significant differences (p-value < 0.05) were found between the 12s window and the majority of window sizes evaluated, with the exception of the 8, 9, 10 and 11 second sizes; however, the standard deviation decreased as the window size increased. Therefore, we set the 12s sliding window as the optimal size for EEG feature extraction regarding positive and negative emotional valence classification.

Once that the comparison between the different window sizes had reveal the most appropriate time window for feature extraction, we have taken a deeper look into the 12 seconds time window, in order to find statistical differences between the classifiers' performance. The statistical analysis reveal three different groups depending on the reached classification performance. The groups were significantly different from each other, but such differences did not exist between components of the same group (Fig. 4). The first group, which exhibited worst performance in comparison with the others, was formed by the GB and GNB classifiers. Second group, was formed by the next classifiers: LDA, QDA, l-SVM, nonl-SVM and RF. Finally, the third group was formed by the same classifiers as the second group but changing the LDA classifier for the KNN. Moreover, we could point out that, although there were no significant differences between them and the rest of sorters of the same statistical group, the classifiers that exhibited the best results with lowest variability between subjects were QDA and KNN. We will focus on these two classifiers for further analysis. The mean classification accuracies reached by the QDA and KNN classifiers at the 12s time window were 99.69%(std 0.48) and 99.75%(std 0.78), respectively.
Frontal

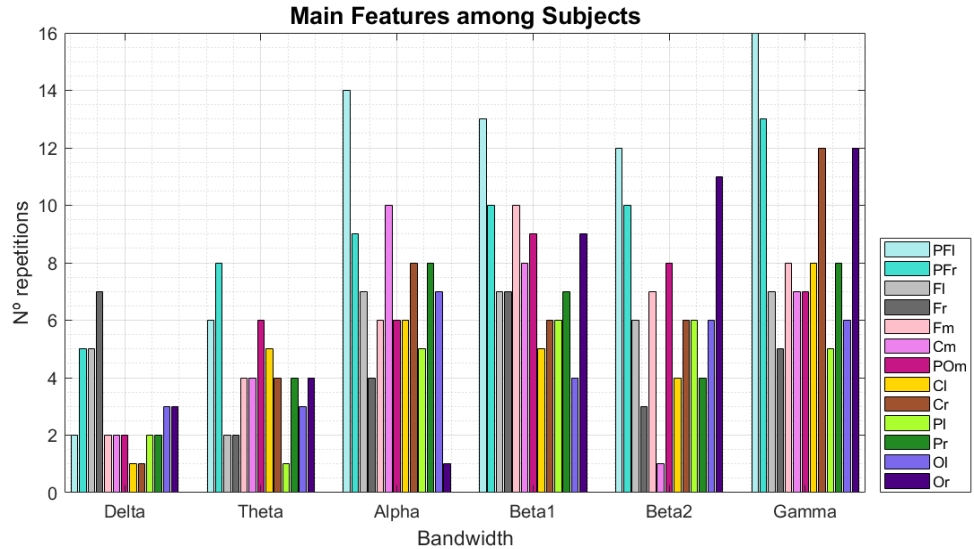

Figure 5. Distribution of the 20 selected frequency-location features of the 24 subjects.

### 3.3. Relevant Frequency-location Variables

Subsequent to the selection of the suitable sliding window size and the classifier, it is necessary to deal with the huge amount of data and therefore useless information that enters the classifier, acting as noise. The dimensionality reduction is not only useful in order to decrease the signal to noise ratio and improve the bias-variance trade off, but also is important to reduce the computational cost for real-time applications. In that way, we performed RFE analysis at the SD model with the 12s sliding window, so as to select those frequency-location variables that carry the relevant information to allow positive and negative emotion recognition (Fig. 5). It can be observed that the most repeated features accumulate less in the low frequencies, delta and theta. With respect to the cortical regions that present more features shared between subjects, we can highlight the prefrontal cortex. In sequence, the 20 most repeated features among subjects with the PR criterion, were finally chosen for final classification stage (Fig. 6). Of the 20 features proposed to form the final classification model, we can highlight B1 band along midline, gamma band throughout most of the scalp, high frequencies at the right occipital region and that the prefrontal area seems to be key on almost every frequency. On the other hand, same procedure was repeated for the SR label criterion, the 20 most repeated features turned out to be the same as in the PR criterion with the exception of 6 of them. The pairs founded at the PR criterion but not in the SR were B1-Cm, B1-POm, B1-Or, B2-POm, Gamma-Fm and Gamma-Pr. Conversely, the features resulted as relevant at the SR but not at the PR were Theta-PFr, Alpha-Pr, B2-Fr, Gamma-Fl, Gamma-POm and Gamma-Pl. If these differences are critical or not will be confirmed in the next section.

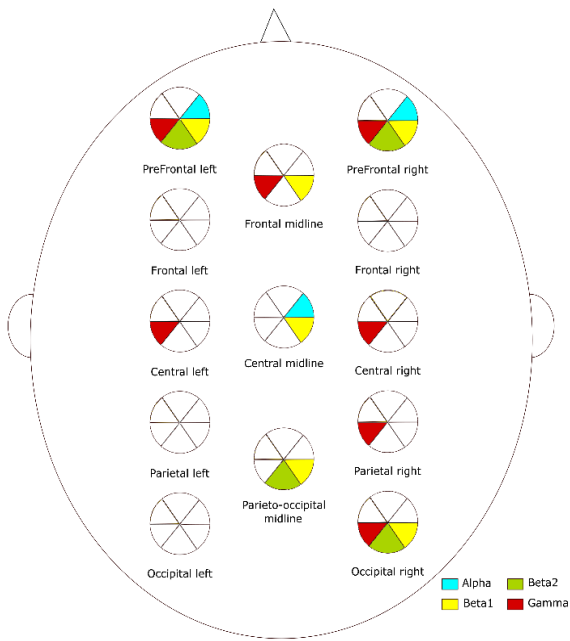

Figure 6. The 20 most repeated frequency-location features among subjects at the 12s sliding window, proposed for final classification model.

### 3.4. Final Classification

After the selection of the potential groups of frequency-location variables, we performed a new classification through the optimal selected classifiers, QDA and KNN, at the 12s sliding window. We have performed an SD classification and calculated the mean f1 scores for all subjects with the different groups of features (Table 1). The group that has obtained the best classification results is the one in which the initial 78 features were used. However, although its mean was slightly higher, there were no significant differences compared to the results obtained by the groups formed by 20 features, with common SD-PR features, and individual SD-PR for the QDA classifier; and with the same two groups and the individual SD-SR features group for the KNN classifier. However, the three groups of 20 features showed no differences within them or with the SI-SA group of variables, this occurred in the two classifiers. The results indicate four possible conclusions. First, reducing the dimensionality to 20 frequency-location pairs gives a result similar to that when classifying taking into account all frequency bands and brain locations. Second, classification performance is affected by using fewer than 20 features. Third, there are no differences between using custom variables for each subject or using the common variables, both using the SD-RFE approximation and the SI-SA approximation, although with the common SD-RFE features one is closer to the classification result achieved with all features. Finally, no significant differences have been found between the different trial labelling criteria, however, SD-PR classification obtained slightly better results than the SD-SR; so we can suggest in a preliminary way that subjective scores are not relevant when classifying emotions on the valence scale. Moreover, the KNN classifier obtained better results than the QDA and the differences between them were statistically relevant (p-value < 0.05) for the conditions of all features, 5 features and 20 common SD-PR features. In figure 7, the differences in the performance of each classifier can be better observed and compared.

|  | All features | 5 common features SD-PR | 10 common features SD-PR | 15 common features SD-PR | 20 common features SD-PR | 20 individual features SD-PR | 20 individual features SD-SR | SI-SA features |
|---|---|---|---|---|---|---|---|---|
| QDA | 0.997 (±0.005) | 0.786 (±0.067) | 0.886 (±0.048) | 0.949 (±0.035) | 0.98 (±0.014) | 0.981 (±0.015) | 0.979 (±0.014) | 0.975 (±0.021) |
| KNN | 0.997 (±0.008) | 0.922 (±0.056) | 0.968 (±0.028) | 0.985 (±0.014) | 0.989 (±0.013) | 0.993 (±0.01) | 0.992 (±0.011) | 0.988 (±0.012) |

Table 1. F1 scores for QDA and KNN classifiers at the different groups of potential features

### 3.5. Subject-Independent Approximation

Regarding SI approximation, SA optimization method was implemented at the 12s sliding window. As the optimization was conducted for each of the eight classifiers evaluated, different number of features were selected for each of them. When looking for the most repeated characteristics among the classifiers, we found that 8 of the frequency-location pairs were present in 5 of the 8 classifiers evaluated and one pair in 7 of the 8. The most repeated characteristic was the Delta band in the left parietal region; and the next most repeated ones corresponded to Delta-PFl, Delta-Fm, Theta-PFl, Theta-Cr, Alpha-Pr, Alpha-Ol, B1-Pfl and B1-PFr. This time, the classification stage, was performed on the one hand taking into account the 20 common features extracted from the SD-RFE method and on the other hand, the features resulting from the optimization of each classifier.

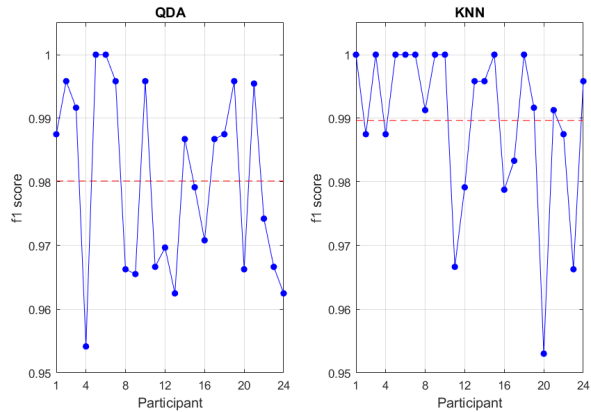

Figure 7. Classification results for the 24 subjects at the 12s sliding window using the 20 common SD-PR features. The red dotted line indicates the mean f1 score.

F1 scores are showed on Table 2. The classification results obtained for the whole population with the 20 common features obtained from the SD-RFE approach were no different from chance. In addition, although the results obtained individually by each classifier, after optimization, were neither significantly different from chance, the averages were slightly better compared to the previous classification approach. The best results, although we insist, not significant, were achieved by the linear classifiers LDA and l-SVM. Given the low performance of the classification, we could highlight the problem of the huge inter-subject variability, and therefore, the difficulty of finding a common computational model that might avoid or at least reduce the training phase of the subjects, prior to any sorting.

| Classifier | 20 common features SD-PR | Features from SA | Nº features SA |
|---|---|---|---|
| GNB | 0.469 (± 0.154) | 0.535 (± 0.177) | 23 |
| LDA | 0.551 (± 0.095) | 0.6 (± 0.097) | 27 |
| QDA | 0.524 (± 0.082) | 0.552 (± 0.083) | 20 |
| l-SVM | 0.569 (± 0.094) | 0.589 (± 0.091) | 35 |
| nonl-SVM | 0.529 (± 0.102) | 0.552 (± 0.093) | 42 |
| GB | 0.459 (± 0.108) | 0.58 (± 0.096) | 23 |
| KNN | 0.522 (± 0.092) | 0.541 (± 0.061) | 25 |
| RF | 0.529 (± 0.01) | 0.579 (± 0.095) | 31 |

Table 2. F1 scores for the eight studied classifiers at the 12s sliding window.

4. **Discussion**

The need to include emotional intelligence skills in machine intelligence is ever more present; new generations are growing hand in hand with technology, and as a consequence, a shift on thinking of new technology development either on social and educational aspects is needed. The range of applications that could be cover by aBCI technology by recognizing emotions in real-time is immense, from clinics, applied to mental and motor disorders, to entertainment markets and business. Accordingly, implementing the formula for the development of a reliable and efficient system capable of real-time emotion recognition that could lead to an economical, accessible and easy-to-use device, non-limited to the hospital or research environment, is one of the main targets on nowadays aBCI. Our results contribute to the delimitation of factors such as the optimal sliding window size, the target frequency-location variables, the appropriate classifier and the stimuli labeling criterion, for real-time implementation of positive and negative emotion recognition based on EEG.

The non-stationarity and non-linearity features of the EEG signal require that the data should be treated with different methodologies beyond the classical Fourier transform [42]. By its ratio between power spectral estimation efficacy and time consuming cost, wavelet transform is a good candidate for real-time EEG based emotion recognition. Authors as Ting et al. [43] have proposed the wavelet packets decomposition method as the most suitable for EEG feature extraction. Based on it for power spectral estimation, our results revealed a rising tendency of improving classification performance by increasing the size of the sliding window, reaching the best classification accuracy at the highest evaluated time of 12s. This fact is in line with the results obtained in our previous work [24], were likewise, the 12s window size turned out to be the suitable trial window size to unmask the electrical patterns that represent emotional processing. However, a plateau was reached starting from the window of 8s. These results suggest that emotions are best recognized and probably encoded at longer time periods, around 8-12 seconds. We have checked only until 12 seconds sliding window based on the work of Candra et al. [19], in this, almost 100% success was reached in the majority of subjects; yet, it would be interesting to evaluate larger window sizes to see if, at some point, 100% success for all subjects is reached, if it is possible to do so, or with the expectation of observing a decrease in performance, which would more certainly indicate the glimpse plateau and therefore delimit the emotional time. Regarding the best classification algorithm, GNB and GB showed the worst results, these classifiers have been used more and have obtained good results in emotional classification when the substrate was an image, such as a facial expression [44] or texts [45], such as descriptions of films or products [46] or even twitter opinions [47]. Contrary, both linear and non-linear SVM, QDA, RF and KNN exhibited the best results in the SD approach, without finding significant differences in their performance. These classifiers are commonly used in the classification of emotions based on EEG, however, there are discrepancies when it comes to choosing the most optimal one. Aishah et al. [48], after comparing various studies using different types of SVM for EEG-based emotion classification, concluded that regardless of the type of classifier, extraction protocol and feature selection, in general, SVMs always perform well, arguing two main properties. First, SVMs have a good performance in generalization problems, i.e., with high-dimensional data and small training sets. And second, their flexibility in selecting a specific kernel makes them able to solve complex linearly inseparable problems. Another study carried out by the authors Zheng et al. [49], in which they evaluated the performance of 10 algorithms in classifying positive and negative emotional videos based on differential entropy values in different frequency bands of the EEG signal, showed that linear regression, l-SVM and RF classifiers obtained the best results. However, they propose that the stacking method, which combines several sorters, is more effective than using individual ones. The complexity of the emotion classification problem also suggests that the data will need a more flexible classifier; in general, non-linear classifiers have shown a slightly higher classification performance than that obtained by linear classifiers in BCI based on EEG [50]. The success rate of a classification is not only influenced by the type of classifier used, but also by the recording equipment, the positioning and number of electrodes, the emotional model studied, the signal pre-processing techniques, the method of feature extraction, and finally, the features selected as classifier input. For all these reasons, the choice of a single method for the classification of emotions based on EEG is a complicated task, which lacks consensus today.

Lotte el al. [51], propose a guide for selecting the classifier depending on its ability to cope with specific feature problems such as the presence of noise and outliers, high dimensionality data sets, non-stationarity of the signals, the time information and the size of the training sets. Thus, they suggest that there is no single classifier to approach BCI classification based on EEG, but should be chosen accordingly to the experimental conditions. For the moment, there is no perfect classifier, a balance in classifier flexibility should be achieved to obtain faithful and proficient classifications, while avoiding failed classification performances, overfitting and data mapping. Based on the results obtained in the present study, we propose the QDA and KNN algorithms as the most appropriate for SD emotion classification, as they presented slightly better average classification results than the rest of the studied classifiers and lower standard deviations, thus proving to be more robust among subjects. However, although both could be used interchangeably, the KNN has obtained a slightly greater success.

In the case of cortical locations, both pre-frontal and central hemispheres, the whole midline and particularly the right parietal-occipital zone were highlighted as the most common informative areas among subjects, on SD classification model. Moreover, the interhemispheric differences presented on parietal-occipital lobe made notable a possible asymmetry pattern over this region. All bandwidths have been reported as been involved in emotional processes by several authors. Delta rhythm has been associated with motivational states reflecting the reward system on prefrontal cortex [52]. Increases in Theta activity on fronto-medial regions have been related with positive valence auditory stimuli [53] [54]. Moreover, it is suggested that Theta band plays an important role on central executive function by integrating affective and cognitive information [55]. In the case of the Alpha bandwidth, it has been observed a desynchronization on posterior scalp regions, representing increased sensory processing [56]; and regarding emotions, frontal Alpha asymmetries have been reported by diverse authors as a function of approach/withdrawal motivational systems [57] [58]. Beta rhythm, more related with the sensory-motor system [59], has been also related with affect over increments of its activity on temporal regions [60], but a general decrease in front of emotional impact stimuli was also reported [61]. Last one, gamma rhythm, is related with the integration of different sensory and non-sensory information [62]. Increased Gamma has been linked to positive valence stimuli [63]; and in particular, frontal gamma power has been related to emotional impact stimuli [62]. In this context, it is interesting that in our results, low frequencies, Delta and Theta, did not appear as relevant in the encoding of emotional valence. Contrary, Gamma band seems to be key in the process, showing an expanded distribution almost over the entire cortical surface. Differences in frequency-location pairs reflecting emotion processing may also vary between studies due to factors such as electrode positioning; type of stimulus (visual, auditory, audiovisual, etc.); the parameters used for feature extraction, as the size of the time window; and the model of emotion studied, since for example it is not the same to focus on the search for discrete emotions such as fear or joy, than to focus on the emotional dimensions such as the pleasure/displeasure scale or what is the same, the positive/negative scale. Once again making clear the lack and need for a common protocol.

On the other hand, a SI approximation would be more desirable for the study of the relevant features that encode emotions, since by using the information of all the subjects at the time, the common characteristics between them become more evident and the noise coming from the inter-subject variability is eliminated. This will make BCI systems more accessible, population reliable and fastest due to it would be possible to avoid the subject training session and classifier adaptation. At the present study, the resulting regions and frequencies obtained with the SI approximation, did not coincide in their majority with the features resulting from the SD approximation; highlighting the Delta and Theta low frequencies, the Alpha rhythm in posterior areas, and coinciding only in the B1 band in the prefrontal lobe. However, these results are not conclusive because of the low percentage of success achieved in classification, therefore, evidencing the widespread and persecuted problem of the SI approach in neural data sorting. Several authors coincide on the idea that the variability is essential for the optimal function and adaptability to the environment of the nervous system [64] [65] [66]. Brain variability is important to perform faster, more accurate and with more capacity of adaptation across tasks or trials, therefore, mean-based brain measures do not represent the actual processes of brain function [66]. However, not all the regions of the brain are benefit by variability [66]; and it has been showed that poor-performance, older brains presented less variability than the youngest, good-performance brains, suggesting some evidence for the dedifferentiation theory [67] [68], and that signal variability follows an inverted-U trend from childhood to old age [69]. Moreover, individual brain signals are unstable between subjects and also, intra-subject sessions because of varying impedances, misalignment of electrode placement, user fatigue, different mental and physical conditions, attention level, etc. [70] [71] [69]. In this respect, several authors have proposed methodologies to deal with the problem of inter-subject and intra-subject session-to-session variability by looking for transfer learning patterns in neuroimaging data [71] [72] [73]. For example, Chai et al. [74] have proposed a new method known as the subspace alignment auto-encoder (SAAE), for EEG data emotion recognition, focus on the distribution discrepancy between subjects. This model exhibited a better performance than state-of-the-art methods, and decreased the discrepancy and reduced performance degradation inter-individuals and inter-sessions. It also pointed out even more the difficulty of classifying EEG patterns across subjects, based on traditional classifiers. In conjunction with Chai et al., Morioka et al. [72] claimed for the accommodation of the variability in brain signals across subjects and intra-subject sessions, because such variability degrades model or classifier performance. Therefore, brain variability is not something we can get rid of or ignore, but more progress needs to be made to understand it and study the factors that generate it, before we can deal with a common formula for the classification of emotions or any BCI application, following an SI approach. In addition, it would be optimal for future studies to control experimental conditions such as the time of day in which stimulation is performed, the ingestion of psychoactive substances such as coffee, the emotional state and stress of the subjects, as well as establish more delimited age ranges.

Finally, we were interested in one more factor of the classification, the stimuli label criterion. As all people express their emotions differently, it is not an easy task to model human emotions and this fact could contribute to the inter-subject variability, and therefore impair classification performance. Our results showed no significantly relevant differences between the PR and the SR criteria, although PR performance was slightly better than the SR, meaning in a preliminary way that both criteria could be use without distinction on EEG based emotion recognition systems. In this context, PR criterion would be more useful in order to achieve easier real-time BCI systems, since individual ratings could be omitted, and therefore, the preparation time and complexity of the process could be reduced. Although more specific studies of this particular parameter with larger sample size would be necessary, the preliminary results pointed out that the subjective assessment of the valence dimension, and probably also those of arousal and dominance, do not faithfully represent the underlying emotional process; in other words, as the ratings are emitted a posteriori of the emotion, they may be contaminated by the conscious rational thought that evaluates the emotional process based on previous experiences and cultural stereotypes, instead of simply giving a value to the lived experience [75]. The best example that reflects this hypothesis is the case of videos whose content was horror, participants experienced fear (they became tense, and even some were scared), but, nevertheless, gave a positive valence score. We believe that it is key here to distinguish both processes, emotion such as fear, whose valence is negative, and the rationalization of emotion, the conscious evaluation of the emotional experience that can modify the valuation of the lived experience, but cannot change the emotion experienced during the visualization of the video. In this way, we want to emphasize the importance of labeling stimuli based on common criteria of their content and not only on the individual subjective score.

The great amount of algorithms that could be applied for BCI, generates endless combinations and therefore, comparison between studies become an arduous task. It is important to uncover a robust and reproducible protocol for EEG based emotion recognition and classification. Furthermore, it would be interesting to study the implication of each frequency-location ensemble on each type of emotion experimentation, through cortical asymmetries and coherence analysis. Lately, it is important to take into account the information that could be obtained from other physiological non-brain signals at the time of emotion classification. It would be beneficial to assess the contribution of several methodologies, as heart rate variability [76] or galvanic skin response [77], in order to generate a more complete and efficient model of aBCI.

## 5. Conclusions

Specify the optimal parameters to perform a real-time aBCI system, capable of recognizing emotions based on EEG, is crucial in order to compare studies among authors and obtain a robust and accessible system. On this sense, we have tried to provide some consensus, at least in the SD approach, setting parameters such as the optimal time base for feature extraction, the classifier with the best performance when classifying positive and negative emotions, the set of informative frequency bands and cortical regions that encode this emotional valence process, and the criterion for the labelling of stimuli. Further efforts should be done now for testing, in a real-time application, the proposed model for positive and negative emotion recognition and continue to study the decoding of the SI approach.

**ACKNOWLEDGMENT**
This work was supported in part by a grant from the Ministry of Education of Spain (FPU grant AP2013/01842), the Spanish National Research Program (MAT2015-69967-C3-1), the Spanish Blind Organization (ONCE) and the Seneca Foundation - Agency of Science and Technology of the Region of Murcia.

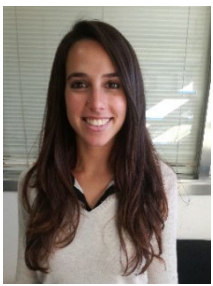
**Jennifer Sorinas** received her Degree in Biomedical Science from Universitat de Lleida, in 2013, and the Master Degree in Neuroscience from the Universidad Miguel Hernandez de Elche, Spain, in 2014, working on the development of a real-time functional near infrared system. She is currently pursuing her PhD degree in neural engineering at the same university, granted by the Government of Spain. She has performed granted research stays at Pittsburgh University – Carnegie Mellon University, USA; and at New York University, USA. Especially, she is interested in the neural substrates that codify and process emotional information.

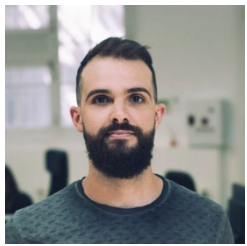
**Juan C. Fernandez-Troyano** received his Degree in Mathematics in 2014 from Unifversidad de Cantabria, then he coursed a Master Degree in Image and Signal Processing from University of Salzbur. He is currently working as Data Scientist in a software company. His main interest are machine learning, feature extraction and data processing.

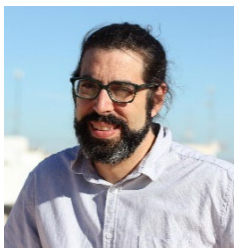
**Mikel Val-Calvo** received his Degree in Computer Engineering in 2016, then he obtained a Master Degree in advanced I.A. from Universidad Nacional a Distancia (UNED), Spain in 2017. He is currently doing his Ph.D. at UNED and he is affiliated to the Department of Electronics and Computer technology at the University of Cartagena. His primary research interests are in artificial intelligence and robotics. More specific interests include machine learning, signal processing and affective computing.

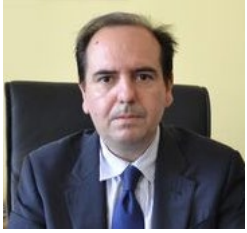

**Jose Manuel Ferrández** received the PhD degree in Informatics with honors from the Technical University of Madrid, Spain, in 1997. He made his postdoctoral stay in the Department of Neurobiology, at the University of Oldenburg, Germany. He is the director of the Electronic Design and Signal Processing Techniques Group and Vice-rector for Internationalization and Development Cooperation, at the University of Cartagena, Spain. He is an evaluator and advisor to the European Commission, in the Future Emerging Technologies Neuroinformatics program of H2020.

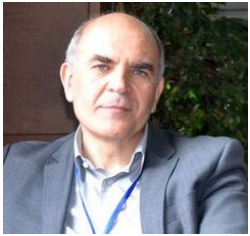

**Eduardo Fernandez** received the PhD degree from the University of Alicante, Spain, in Neuroscience with honors in 1990. He is Chairman of the Department of Histology and Anatomy in the University Miguel Hernández (Spain) and Director of the Neuroengineering and Neuroprosthesis Unit at the Bioengineering Institute at the same university. His research interest is in developing solutions to the problems raised by interfacing the human nervous system and on this basis develop a two-way direct communication with neurons and ensembles of neurons. He is actively working on the development of neuroprostheses and brain-machine interfaces.